\documentclass[showpacs,aps,preprint]{revtex4}
\usepackage{graphicx}
\usepackage{dcolumn}
\usepackage{amsfonts}
\usepackage{amsmath}

\newcommand{\ep}{\epsilon}

\newcommand{\lraw}{\longrightarrow}

\newcommand{\pa}{\partial}
\newcommand{\td}{\tilde}

\newcommand{\beq}[1]{\begin{eqnarray}\label{#1}}
\newcommand{\eeq}{\end{eqnarray}}

\newcommand{\llraw}{-\!-\!\!\!\lraw}

\newcommand{\CN}[1]{{\cal N}=#1}
\newcommand{\nn}{\nonumber}

\begin{document}
\title{Spinning Strings on Deformed $AdS_5\times T^{1,1}$ with NS
$B$-field}
\author{Xiao-Jun Wang}
\email{wangxj@ustc.edu.cn}
\affiliation{\centerline{Interdisciplinary Center for Theoretical
Study} \centerline{University of Science and Technology of China}
\centerline{AnHui, HeFei 230026, China}}

\begin{abstract}
We study classical spinning closed string configuration on
logarithmically deformed $AdS_5\times T^{1,1}$ background with
non-trivial Neveu-Schwarz $B$-field in which IIB string theory is
dual to a non-conformal $\CN{1}$ $SU(N+M)\times SU(N)$ gauge
theory. The integrability on original $AdS_5\times T^{1,1}$
background are significantly reduced by $B$-field. We find several
spinning string solutions with two different ansatzs. Solutions
for point-like strings and few circular strings are explicitly
obtained. Folded spinning string solutions along radial direction
are shown to be allowed in this background. These solutions
exhibit novel properties and bring some challenges to understand
them from dual quantum field theory.
\end{abstract}
 \preprint{USTC-ICTS-05-01} \maketitle

\section{Introduction}

Recent two years the test of AdS/CFT correspondence was received
significant development due to discovery of integrability both in
string theory on $AdS_5\times S^5$ background and in $\CN{4}$
super Yang-Mills theory (SYM). It allows us to test AdS/CFT beyond
BPS sector. The integrability at the string theory side was
indicated first by discovery that the IIB superstring theory is
exactly solvable at pp-wave background\cite{Metsaev02,BMN02}, and
later by finding of various semiclassical spinning solutions of
string sigma model on $AdS_5\times S^5$
background\cite{GKP02,FT02,Minahan02,AFRT03} (see the
reviews\cite{Tse03,Tse04} and references therein). On the other
hand, at the gauge theory side the integrability is exhibited as
identification between dilatation operator of $\CN{4}$ SYM and
Hamiltonian of a integrable spin chain\cite{MZ02,Beisert03} (see
the review\cite{Beisert04} and references therein). Following
those developments the test was naturally extended into general
gauge/string duality, with less
supersymmetries\cite{WY03,Ideguchi04,VT04} and without conformal
symmetry\cite{NCa,T11,Kruc03,NCb}.

The purpose of this present paper is to study the spinning closed
string configurations on the Klebanov-Tseytlin (KT)
background\cite{KT00} which is logarithmic deformation of
$AdS_5\times T^{1,1}$ with non-trivial Neveu-Schwarz $B$-field and
is conjectured to be dual to a non-conformal $\CN{1}$
$SU(N+M)\times SU(N)$ gauge theory$^1$\footnotetext[1]{It is worth
to point out that various spinning string solutions have been
found in $AdS_5\times T^{1,1}$ background\cite{T11}. Those
solutions, however, were extracted without consideration of the NS
$B$-field and logarithmic deformation of background, or are
presented on $AdS_5$ part of background where the NS $B$-field
plays no role.}. This study is expected to enable us to precisely
test general gauge/string duality beyond supergravity level, like
those for AdS/CFT correspondence. One motivation to choose the KT
background is that this background simple enough, but possesses
some novel properties comparing with those pure Ramond-Ramond
backgrounds: The NS $B$-field directly couples with world-sheet
scalar fields. Hence a non-constant $B$-field enters equations of
motion. The backreaction of $B$-field induces a logarithmic
deformation to $AdS_5\times T^{1,1}$ background. The resulted
geometry is no longer to be simply product of $AdS_5$ and
$T^{1,1}$. Rather, there are warped factors both for $AdS_5$ and
$T^{1,1}$ metric. In other words, string motion in $AdS_5$ and in
$T^{1,1}$ are no longer to be decoupled. The above two properties
must significantly alter the integrability as well as the spinning
string configurations on original $AdS_5\times T^{1,1}$
background.

In the present paper we focus our attention on the multi-spin
solutions that describes the closed strings rotating on $T^{1,1}$
part only. It according to AdS/CFT dictionary corresponds to the
composite operators consisting of bi-fundamental scalar fields of
a $\CN{1}$ $SU(N+M)\times SU(N)$ gauge theory. Here the isometry
of $T^{1,1}$, $SU(2)\times SU(2)\times U(1)$, is dual to a global
symmetry in SYM. Hence our another motivation to study
semiclassical string configurations on the KT background is that
the dictionary of duality about this background is much clearer
than one on other confining backgrounds\cite{NCb}. In other words,
even though various spinning string solutions were obtained on
those confining backgrounds, it is hard to exactly construct the
dual operators of the most of spinning string solutions and check
gauge/string duality directly. For example, we even can not
exactly know what quantum numbers in SYM correspond to conserved
charges associated with strings rotating on transverse space.

The above two advantages should admit us to study the
non-conformal properties of gauge theories from perspective of
semiclassical solution of dual string theory. It is beyond those
extracted from supergravity analysis. One typical property was
known as the absence of folded string configuration along radial
direction in confining background\cite{NCb}. To be precise, the
strings spreading along radial direction of transverse space are
no longer to satisfy the periodic condition. This conclusion
follows from the well-known radius/energy-scale
correspondence\cite{PP99,WH02}. In other words, the physical
parameters like coupling constants are changed with flow along
radial direction. It makes that the semiclassical energy of
strings as functions of conserved charges $J$ as well as 't Hooft
coupling $\lambda$ to be ambiguous. In this present paper,
however, we shall show that the folded string along the radial
direction is allowed by the KT background. The key point is that a
combination of two coupling constants of dual SYM, $g^{-2}\sim
g_1^{-2}+g_2^{-2}$, does not slide under renormalization group.
Hence the semiclassical energy as function of $J$ and $g^2$ is
well-defined.

One of the essential point to test the gauge/string correspondence
is to explore the energy/charge relations for various
semiclassical string configurations. This relation at large charge
$J$ limit usually admits a regular expansion in positive powers of
$\lambda/J^2$ with $\lambda$ the 't Hooft coupling of dual SYM.
The semiclassical expansion of string sigma model is valid at
large $\lambda$ limit. Hence we actually need take double limit,
$\lambda\to\infty,\;J\to\infty$, while $\lambda/J^2$ fixed. In
this limit the classical energy of semiclassical string solution
in any $AdS_p\times S^q$ space goes as linear function of $J$,
$E=J+...$ with appropriate normalization. This linear
behavior\cite{VLS94} is a consequence of the constant curvature of
$AdS$ space. In our case, however, we expect that the linear
behavior is breakdown since no matter $AdS$ part or $T^{1,1}$ are
no longer to possess constant curvature due the logarithmic
deformation. Indeed we find that the relation may be modified to
$E=E_0+F(J)$ with $E_0>0$ when string spreads along the radial
direction since the curvature of deformed $T^{1,1}$ varies along
the radius. While the linear behavior is kept for point-like
(BMN-like) and circular configurations since they locate at a
fixed position on radial direction. Moreover, we find that the
point-like solutions is only presented at a special point on
radial coordinate of deformed $AdS_5$. While the circular
solutions are allowed only in a finite region on radial coordinate
of deformed $AdS_5$. This phenomena should somehow exhibit
non-conformal properties of dual gauge theory.

This paper is organized as follows: In section 2 we briefly review
the KT background and some basic points of string sigma model on
this background. In section 3 we study spinning string solutions
with usual ansatz. We argue that some simpler multi-spin circular
solutions are absent in this ansatz. We also show that the folded
string configuration is allowed in the KT background, but it
possesses infinite energy. In section 4 various spinning string
solutions are constructed with another ansatz. A brief summary is
devoted in section 5 and some further discussions and open
questions are also included in this section.

\section{String sigma model on the Klebanov-Tseytlin background}

\subsection{The Klebanov-Tseytlin background}

The Klebanov-Tsetylin geometry\cite{KT00} is a singular background
produced by placing $N$ D3-branes and $M$ fractional D3-branes at
the apex of conifold. The angular part of conifold is a cone whose
base is a coset space $T^{1,1}=(SU(2)\times SU(2))/U(1)$, with
topology $S^2\times S^3$ and symmetry group $SU(2)\times
SU(2)\times U(1)$. The 10-d metric of KT geometry has the
structure of a ``warped product'' of $\mathbb{R}^{3,1}$ and the
conifold:
\beq{2.1}
ds_{10}^2=h^{-1/2}(r)\eta_{\mu\nu}dx^\mu dx^\nu +
h^{1/2}(r)(dr^2+r^2g_{ij}dy^idy^j),
\eeq
where $\mu,\nu=0,1,2,3$ and $i,j=4,...,9$, $g_{ij}$ denotes the
Einstein metric on $T^{1,1}$,
\beq{2.2}g_{ij}dy^idy^j=\frac{1}{6}\sum_{i=1}^2(d\theta_i^2
+\sin^2\theta_id\phi_i^2)+\frac{1}{9}(d\psi+\cos\theta_1d\phi_1
+\cos\theta_2d\phi_2)^2.
\eeq
The insertion of fractional D3-branes involves $M$ units of the
R-R 3-form flux $F_3$ as well as radial dependent NS 3-form flux
$H_3=dB_2$:
\beq{2.3}F_3=\frac{M\alpha'}{2}\omega_3,\hspace{1in}
B_2=\frac{3g_sM\alpha'}{2}\omega_2\ln{(r/r_0)},
\eeq
where
\beq{2.4}\omega_2&=&\frac{1}{2}(\sin\theta_1d\theta_1\wedge d\phi_1-
\sin\theta_2d\theta_2\wedge d\phi_2), \nn \\
\omega_3&=&(d\psi+\cos\theta_1d\phi_1 +\cos\theta_2d\phi_2)
\wedge\omega_2.
\eeq
The warped factor $h(r)$ is determined by solving the Einstein
equation in presence of 3-form flux,
\beq{2.5}h(r)=\frac{R^4}{r^4}[1+a\ln{(r/r_0)}],
\eeq
where
\beq{2.6}R^4=\frac{27}{4}\pi\alpha'^2(g_sN+\frac{3}{8\pi}g_s^2M^2),
\hspace{0.6in} a=\frac{12g_sM^2}{8\pi N+3g_s M^2}.
\eeq
Hence this geometry is a logarithmic deformation of $AdS_5\times
T^{1,1}$. It has a naked singularity at $r=r_s$ where $h(r_s)=0$
and approaches to $AdS_5\times T^{1,1}$ in vicinity of the UV
scale $r=r_0$. An important feature of this background is that the
5-form flux acquires a radial dependence,
\beq{2.7}\td{F}_5={\cal F}_5+\star{\cal F}_5,\hspace{1in}
{\cal F}_5\propto N_{\rm eff}=N+\frac{3}{2\pi}g_sM^2\ln{(r/r_0)}.
\eeq
The reduction of 5-form flux along radius is known as RG cascade
in dual gauge theory. In particular, $\td{F}_5$ vanishes at
$r=\td{r}_s=r_0\exp{(\frac{1}{4}-\frac{1}{a})}$ where
$h(r_s)\propto 3g_sM^2/(8\pi N+3g_sM^2) > 0$. It is well-known
that the naked singularity can be removed by blow-up of $S^3$ at
apex of $T^{1,1}$, it is Klebanov-Strassler solution\cite{KS00}.
In this present paper, however, for the sake of simpleness we
focus on classical string configuration in the region $r>\td{r}_s$
which does not touch the singularity. Hence we still work on the
KT background.

IIB superstring theory on KT background is conjectured to be dual
to the $\CN{1}$ SYM with $SU(N+M)\times SU(N)$ gauge symmetry. The
matter sector of the model consists of two chiral superfields
$A_1,A_2$ in the $({\bf N+M},\overline{\bf N})$ representation and
two fields $B_1,B_2$ in the $(\overline{\bf N+M},{\bf N})$
representation. The model has a $SU(2)\times SU(2)\times U(1)$
global symmetry and consequently has three $U(1)$ charges for
fields in matter sector. The later corresponds to translation
invariance along $\phi_1,\phi_2$ and $\psi$ in KT geometry. The
quantum numbers$^2$\footnotetext[2]{It should be distinguished the
$U(1)$ charges $U(1)_A$ and $U(1)_B$ in table 1 from those present
in ref.\cite{KS00}. The charges $U(1)_A$ and $U(1)_B$ in our paper
are associated with global $SU(2)\times SU(2)$ symmetry. However,
they in \cite{KS00} are associated with unbroken gauge group
$U(1)^N$.} of the matter under the global symmetries are given in
table 1.

\begin{table}[hptb]
\setlength{\tabcolsep}{0.2in}
 \begin{tabular}{c|ccccc}
 \hline
 & $SU(2)$ & $SU(2)$ & $U(1)_A$ & $U(1)_B$ & $U(1)_R$ \\ \hline
$(A_1,A_2)$ & $\bf{2}$ & $\bf{1}$ & $(1,-1)$ & 0 & $\frac{1}{2}$ \\
$(B_1,B_2)$ & $\bf{1}$ & $\bf{2}$ & 0 & $(1,-1)$ & $\frac{1}{2}$ \\
\hline
 \end{tabular}
\centering
 \begin{minipage}{4in}
 \caption{The quantum numbers of matter under global
 $SU(2)\times SU(2)\times U(1)$ symmetry.}
 \end{minipage}
\end{table}

\subsection{String sigma model on KT background}

IIB superstrings on KT background can be described by a
Green-Schwarz action whose bosonic part is a non-linear sigma
model on string world-sheet,
\beq{2.8}I=-\frac{1}{2\pi\alpha'}\int
d\tau d\sigma\sqrt{-g} \{g^{ab}\pa_aX^M\pa_bX^NG_{MN}(X)
+\ep^{ab}\pa_aX^M\pa_bX^NB_{MN}(X)\},
\eeq
where $g^{ab}$ is the world-sheet metric, $\ep^{ab}/\sqrt{-g}$ is
the world-sheet anti-symmetric tensor, $G_{MN}$ and $B_{MN}$ are
target space metric and anti-symmetric tensor respectively. The
world-sheet symmetries of invariance of reparametrization and of
Weyl scaling allow us to take conformal gauge, namely
$g_{ab}=\eta_{ab},\;\ep^{01}=1$. The classical equations of motion
that follows from the above action in conformal gauge read off
\beq{2.9}\pa_a\pa^aX^P+\Gamma^P_{MN}\pa_aX^M\pa^aX^N
-\frac{1}{2}G^{PQ}H_{QMN}\ep^{ab}\pa_aX^M\pa_bX^N=0,
\eeq
where $\Gamma^P_{MN}$ specifies Christoffel connection of the
target space and $H_{QMN}$ is NS 3-form flux. Moreover, the
action~(\ref{2.8}) is to be supplemented with the usual conformal
gauge constraints expression the vanishing of the total 2-d
energy-momentum tensor
\beq{2.10}\left\{\begin{array}{ll}
&G_{MN}(\dot{X}^M\dot{X}^N+X'^MX'^N)=0, \\
&G_{MN}\dot{X}^MX'^N=0,
\end{array} \right.
\eeq
where the ``dot'' and ``prime'' denote differential to $\tau$ and
$\sigma$ respectively.

Concerning to KT background~(\ref{2.1}), (\ref{2.3}), it is useful
to explicitly write equations of motion for $x_\mu$ and $r$,
\beq{2.11}
&&\pa_a\pa^ax^\mu-\frac{h'}{2h}\pa_ax^\mu\pa^ar=0,
\nn \\
&& \pa_a\pa^ar+\frac{h'}{4h}\pa_ar\pa^ar+\frac{h'}{4h^2}
\pa_ax_\mu\pa^ax^\mu-r(1+\frac{rh'}{4h}) g_{ij}\pa_ay^i\pa^ay^j
\nn \\&&\quad\hspace{0.6in}\quad
-\frac{1}{2}G^{rr}H_{rij}\ep^{ab}\pa_ay^i\pa_by^j=0,
\eeq
where $h'=dh(r)/dr$. Due to the nontrivial $H_{3}$ and
$1+\frac{rh'}{4h}=[4\ln{(r/r_0)}+\frac{4}{a}]^{-1}\neq 0$, the
coordinates of $T^{1,1}$, $y_i$, in general do not decouple with
radial coordinate $r$. We will see that the presence of NS 3-form
field strength in the equation of motion plays significant role to
change the integrability of the simple product $AdS_5\times X_5$
background. This is novel feature of string sigma model on KT
background.

It is well-known that there is residual symmetry on world-sheet
sigma model under transformation $\sigma^+=\tau+\sigma\to
\td{\sigma}^+(\sigma^+),\;\sigma^-=\tau-\sigma\to
\td{\sigma}^-(\sigma^-)$. For flat target space this invariance
allows us to transfer arbitrary non-trivial solutions of
$x^0(\tau)$ or $x^+=x^0+x^1=x^+(\tau)$ to the static gauge
$x^0=\kappa\tau$ or the light-cone gauge $x^+=p^+\tau$. In the
curved background, however, it is in general impossible to achieve
those gauge. For example, in the KT background following from the
first equation in~(\ref{2.11}) it is possible to take static gauge
$x^0=\kappa\tau$ only for those special solutions with $r$
constant. Furthermore, $r$ can not be arbitrary constant due to
the constraint from the second equation of (\ref{2.11}). In simple
product backgrounds such as $AdS_5\times S^5$ or $AdS_5\times
T^{1,1}$ it is determined by $h'/h^2=0\to r=0$ unless $R\to\infty$
with $R$ the radius of $AdS_5$. In the KT background, however, the
value of $r$ depends on motion of string inside $T^{1,1}$. More
generally, we may try to find a class of special solutions in
which $r$ depends on $\sigma$ only and satisfies the period
boundary condition, i.e., $r(\sigma+2\pi)=r(\sigma)$. A special
solution for $x^0$ following from this ansatz is
\beq{2.12}x^0=\kappa\tau+\beta(\sigma),\hspace{1in}
\beta'(\sigma)=c_0\sqrt{h(r(\sigma))},
\eeq
with $c_0$ constant and boundary condition
$\beta(\sigma+2\pi)=\beta(\sigma)$.

In this present paper, we focus our attention on those
configurations describing closed strings rotating on $T^{1,1}$.
Hence we simply set $x^i={\rm constant},\; (i=1,2,3)$.

\subsection{Conserved charges}

The symmetry of $T^{1,1}$, $SU(2)\times SU(2)\times U(1)$, admits
three conserved charges. They are angular momenta corresponding to
strings rotating along $\phi_1,\phi_2$ and $\psi$ directions:
\beq{2.13}
J_A=P_{\phi_1}&=&\frac{1}{6\pi\alpha'}\int_0^{2\pi}d\sigma
r^2\sqrt{h(r)}\{(1-\frac{1}{3}\cos^2{\theta_1})\dot{\phi}_1
+\frac{2}{3}\cos{\theta_1}(\dot{\psi}+\cos{\theta_2}\dot{\phi}_2)\}
 \nn \\
&&-\frac{3g_sM}{8\pi}\int_0^{2\pi}d\sigma\;\theta_1'\ln{(r/r_0)}
\sin{\theta_1}, \nn \\
J_B=P_{\phi_2}&=&\frac{1}{6\pi\alpha'}\int_0^{2\pi}d\sigma
r^2\sqrt{h(r)}\{(1-\frac{1}{3}\cos^2{\theta_2})\dot{\phi}_2
+\frac{2}{3}\cos{\theta_2}(\dot{\psi}+\cos{\theta_1}\dot{\phi}_1)\}
 \nn \\
&& +\frac{3g_sM}{8\pi}\int_0^{2\pi}d\sigma\;\theta_2'\ln{(r/r_0)}
\sin{\theta_2}, \nn \\
J_R=P_\psi &=&\frac{1}{9\pi\alpha'}\int_0^{2\pi}d\sigma
r^2\sqrt{h(r)}(\dot{\psi}+\cos{\theta_1}\dot{\phi}_1+
\cos{\theta_2}\dot{\phi}_2).
\eeq

In addition, the classical energy is associated with translation
invariance along $x^0$. For special solutions like (\ref{2.12}),
its explicit expression is
\beq{2.14}
E=P_0=\frac{\kappa}{\pi\alpha'}\int_0^{2\pi}d\sigma h^{-1/2}(r).
\eeq

\section{Type A solutions}
\setcounter{equation}{0}

Inserting KT metric~(\ref{2.1}) and NS 3-form field~(\ref{2.2})
into equations of motion~(\ref{2.9}) we obtain some highly
involved non-linear equations. To solve them some ansatz are
necessary. Here the word ``type A'' just denotes that in this
section we study the various special solutions with ansatz
$\dot{\theta}_i=\phi_i'=\psi'=0,\;(i=1,2)$. Then equations of
motion of $\phi_i$ and $\psi$ reduce to
\beq{3.1}\ddot{\phi}_i=\ddot{\psi}=0\quad\Rightarrow
\quad\left\{\begin{array}{cc} \phi_i=\omega_i\tau,&\quad\quad
i=1,2 \\ \psi=\omega\tau. & \end{array}\right.
\eeq
Furthermore, the second constraint in (\ref{2.10}) implies
$\beta(\sigma)\sim{\rm constant}$. While another conformal
constraint reduces to
\beq{3.2}
\frac{k^2}{r^2h(r)}=\frac{r'^2}{r^2}+
\frac{1}{6}\sum_{i=1}^2(\theta_i'^2+ \omega_i^2\sin^2\theta_i)
+\frac{1}{9}(\omega+\omega_1\cos\theta_1+\omega_2\cos\theta_2)^2.
\eeq
Using this constraint in equations of $r$ and $\theta_i$ we have
\beq{3.4}
&&\frac{r''}{r}-\frac{r'^2}{r^2}-\frac{\kappa^2}{r^2h}
(1-\frac{a/2}{1+a\ln{(r/r_0)}})-\frac{a(\theta_1'^2+\theta_2'^2)}
{12(1+a\ln{(r/r_0)})} \nn \\
&&\quad\hspace{1in}\quad +\frac{R^2\sqrt{a}}{3r\sqrt{2h}}
(\omega_1\sin\theta_1\;\theta_1'-\omega_2\sin\theta_2\;\theta_2')
=0, \nn \\
&&\theta_1''+\frac{a\;\theta_1'}{1+a\ln{(r/r_0)}}\frac{r'}{2r}
-\frac{\omega_1\sin\theta_1}{3}(2\omega+2\omega_2\cos\theta_2
-\omega_1\cos\theta_1
+\frac{3R^2\sqrt{a}r'}{r^3\sqrt{2h}})=0, \nn \\
&&\theta_2''+\frac{a\;\theta_2'}{1+a\ln{(r/r_0)}}\frac{r'}{2r}
-\frac{\omega_2\sin\theta_2}{3}(2\omega+2\omega_1\cos\theta_1
-\omega_2\cos\theta_2
-\frac{3R^2\sqrt{a}r'}{r^3\sqrt{2h}})=0.
\eeq
For $a=0$ the above equations of motion just describe strings
rotating on $AdS_5\times T^{1,1}$ background where $r$ and
$\theta_i$ decouple. Moreover, to describe closed string
configuration, the following boundary conditions should be
supplemented,
\beq{3.5}\theta_i(\sigma+2\pi)=\theta_i(\sigma)+2m_i\pi,\hspace{1in}
r(\sigma+2\pi)=r(\sigma),
\eeq
with $m_i$ integers.

\subsection{Point-like solution}

The point-like string configurations indicates that none of
coordinates acquires the $\sigma$-dependence. Hence equations of
motion~(\ref{3.4}) reduce
\beq{3.6}
\left\{\begin{array}{lll}r_*=r_0e^{\frac{1}{2}-\frac{1}{a}}&\Rightarrow
\quad\quad & N_{\rm eff}=\frac{3}{8\pi}g_sM^2,  \\
\omega_1=0,&{\rm or}& 2\omega+2\omega_2\cos\theta_2
-\omega_1\cos\theta_1=0, \\
\omega_2=0,&{\rm or}&
2\omega+2\omega_1\cos\theta_1-\omega_2\cos\theta_2=0.
\end{array}
\right.
\eeq
Here we used a fact that $\sin\theta_i=0$ means that the
point-like string locates at north- or south-pole of $S^2$.
Classically it does not make sense to claim the point-like string
rotating along $\phi_i$. Therefore, $\sin\theta_i=0$ implies
$\omega_i=0$. From (\ref{3.6}) we see that value of $r_*$ is
uniquely determined by the number of the background fluxes and
string coupling.
\begin{itemize}
\item[-] Single-spin solution: $\omega_1=\omega_2=0$.
\end{itemize}

The conserved charges corresponding to angular momenta following
from Eq.~(\ref{2.13}) read off
\beq{3.7}J_A=J_B=J_R=J=\frac{\sqrt{2a}}{9\alpha'}R^2\omega.
\eeq
Then using conformal constraint~(\ref{3.2}) we obtain the
classical energy of this configuration
\beq{3.8}E=\frac{2}{3\alpha'}r_*\omega\propto J.
\eeq
According to the table 1, we can see that its dual operators is:
\beq{3.9} {\rm Tr}(A_1B_1)^J.
\eeq
\begin{itemize}
\item[-] Two-spin solution: $\omega_1=0,\omega_2\neq 0$.
\end{itemize}

From Eq.~(\ref{3.6}) it is straightforward to obtain
$\cos\theta_2=2\omega/\omega_2\;\Rightarrow\;|\omega|\leq
|\omega_2|/2$ and conserved charges as follows:
\beq{3.10}
J_B=\frac{\sqrt{2a}}{6\alpha'}R^2\omega_2,\hspace{0.5in}
J_R=\frac{\sqrt{2a}}{3\alpha'}R^2\omega\leq J_B, \hspace{0.5in}
J_A=J_R\cos\theta_1,.
\eeq
Because there is no other constraint to fix the value of
$\cos\theta_1$, the dual operators of this solution is rather
complicated. In particular, the $U(1)_A$ in dual SYM is anomalous
if $\cos\theta_1\neq n/J_R$ with $n,\;J_R$ integers. Finally, the
classical energy reads off
\beq{3.11}E=\frac{2}{\alpha'}r_*\sqrt{\frac{\omega^2}{3}
+\frac{\omega_2^2}{6}}=\frac{\sqrt{6}}{\sqrt{a}R^2}r_*
\sqrt{J_R^2+2J_B^2}.
\eeq
\begin{itemize}
\item[-] Three-spin solution: $\omega_1\neq 0,\omega_2\neq 0$.
\end{itemize}

It is straightforward to obtain the conserved charges
from the solution $\cos\theta_i=-2\omega/\omega_i$,
\beq{3.12}J_A=\frac{\sqrt{2a}}{6\alpha'}R^2\omega_1,\hspace{0.5in}
J_B=\frac{\sqrt{2a}}{6\alpha'}R^2\omega_2,\hspace{0.5in}
J_R=-\frac{\sqrt{2a}}{3\alpha'}R^2\omega.
\eeq
They should be supplemented by the condition
\beq{3.13}|J_R|\leq {\rm min}(|J_A|,|J_B|).
\eeq
The classical energy following from conformal constraint is again
expressed by the above charges,
\beq{3.14}E=\frac{\sqrt{6}}{\sqrt{a}R^2}r_*
\sqrt{2(J_A^2+J_B^2)-J_R^2}.
\eeq
Its dual operators have to be out of holomorphic (or
anti-holomorphic) sector. To see this fact, let us consider an
holomorphic operators consisting of $n_1$ $(A_1B_1)$, $n_2$
$(A_1B_2)$, $n_3$ $(A_2B_1)$ and $n_4$ $(A_2B_2)$. If there are no
global $U(1)$ anomaly, we have
\beq{3.15}J_R=\sum_{i=1}^4n_i,\hspace{0.3in}J_A=n_1+n_2-n_3-n_4,
\hspace{0.3in}J_B=n_1-n_2+n_3-n_4.
\eeq
The above conclusion, however, conflicts with
condition~(\ref{3.13}) unless $n_1=n_2=n_3=0$ which corresponds to
$|\omega_1|=|\omega_2|=2|\omega|\;\Rightarrow\;\sin\theta_i=0$.
According to previous discussion, this result conflicts with
assumption $\omega_i\neq 0$. Therefore, we conclude that the dual
operator of three-spin solution must mixing operators $A_iB_j$
with their Hermitian conjugated partners.

\subsection{Circular solution with $r'=0$}

The circular solutions describe those string configurations in
which at least one of $\theta_i$ acquires $\sigma$-dependence and
satisfies the boundary condition~(\ref{3.5}) with nonzero winding
number $m_i$\cite{FT034}. Here we focus our attention on the
circular solutions with $r'=0$. The solutions with $r'\neq 0$ will
be discussed in the next subsection.
\begin{itemize}
\item[-] single-spin solution: $\omega_1=\omega_2=0$.
\end{itemize}

From equations of $\theta_i$ we have $\theta_i=m_i\sigma$ with
$m_i$ integers. Then the consistency between conformal constraint
~(\ref{3.2}) and the first equation in (\ref{3.4}) implies
\beq{3.16}r_*=r_0\exp{\left(\frac{\omega^2}{3(m_1^2+m_2^2)+2\omega^2}
-\frac{1}{a}\right)}.
\eeq
While the condition $N_{\rm eff}>0$ imposes a constraint
$2\omega^2>3(m_1^2+m_2^2)$.

The angular momenta and classical energy are as follows:
\beq{3.17}
J_A&=&J_B=0,\hspace{0.5in} J_R\equiv
\frac{R^2}{9\alpha'}\sqrt{\frac{2a}{3}}{\cal
J}=\frac{2R^2}{9\alpha'}\omega
\sqrt{\frac{a\omega^2}{3(m_1^2+m_2^2)+2\omega^2}},
\nn \\
E&=&\frac{2r_0}{\alpha'}e^{-1/a}\sqrt{m_1^2+m_2^2+\frac{2}{3}\omega^2}
\exp{\left(\frac{\omega^2}{3(m_1^2+m_2^2)+2\omega^2}\right)} \nn
\\
&&\stackrel{{\cal J}\to\infty}{\llraw}
\frac{2\sqrt{2}r_0}{\alpha'}e^{\frac{1}{2}-\frac{1}{a}}{\cal
J}(1+\frac{m_1^2+m_2^2}{{\cal J}^2}+...) .
\eeq
Hence we again have the regular expansion to classical energy at
large angular momentum limit. Its dual operators have the form
like $${\rm Tr}(A_1B_1)^{J_R/2}(A_2B_2)^{J_R/2}.$$ Otherwise they
should be out of holomorphic sector.

Other single-spin solutions, e.g.,
$\omega=\omega_1=0,\;\omega_2\neq 0$ will be studied in the
following discussions.
\begin{itemize}
\item[-] Two-spin solution: $\omega_1\neq 0,\;\omega_2=0$.
\end{itemize}
The solutions of Eq.~(\ref{3.4}) describing the circular string
configuration is
\beq{3.17s}\theta_2=m\sigma,\hspace{0.3in}
\hspace{0.4in} \cos\theta_1=2\omega/\omega_1={\rm constant}.
\eeq
Then
$$r_*=r_0\exp{\left(\frac{\omega_1^2+2\omega^2}
{2m^2+2\omega_1^2+4\omega^2}-\frac{1}{a}\right)}$$ follows from
the consistency between conformal constraint ~(\ref{3.2}) and the
first equation in (\ref{3.4}). The conversed charges and classical
energy reads off:
\beq{3.17q}
J_A&=&\frac{2R^2}{6\alpha'}\omega_1
\sqrt{\frac{\omega_1^2+2\omega^2} {2m^2+2\omega_1^2+4\omega^2}},
\hspace{0.5in} J_B=0, \nn \\
J_R&=&\frac{2R^2}{3\alpha'}\omega
\sqrt{\frac{\omega_1^2+2\omega^2}
{2m^2+2\omega_1^2+4\omega^2}}\leq J_A,
\nn \\
E&=&\frac{2r_0}{\alpha'}e^{-1/a}\sqrt{m^2+\omega_1^2+2\omega^2}
\exp{\left(\frac{\omega_1^2+2\omega^2}
{2m^2+2\omega_1^2+4\omega^2}\right)} \nn
\\
&&\stackrel{{\cal J}\to\infty}{\llraw}
\frac{2\sqrt{2}r_0}{\alpha'}e^{\frac{1}{2}-\frac{1}{a}}{\cal
J}(1+\frac{m^2}{{\cal J}^2}+...),
\eeq
where ${\cal J}^2\propto \sqrt{2J_A^2+J_R^2}$. Its dual operators
in (anti-)holomorphic sector have the form like $${\rm
Tr}(A_1B_1)^{J_R/2}(A_1B_2)^{J_R/2}.$$ It exists only for
$J_R=J_A$ and even $J_R$. For other choices of charges they are no
longer to be (anti-)holomorphic.

If we set $\omega=0$, the above solution reduces to the
single-spin solution with $\omega=\omega_1=0,\;\omega_2\neq 0$.
Moreover, It should be pointed out that NS $B$-field plays trivial
role in all of the above solutions. In the following we will see
that some simpler multi-spin solutions (such as those found in
$AdS_5\times X^5$ background) are absent when NS $B$-field plays
nontrivial role.

\begin{itemize}
\item[-] Fate of other multi-spin circular solutions with $r'=0$.
\end{itemize}
It should be noticed that the conformal constraint~(\ref{3.2}) is
consistent with the equations of motion of $\theta_i$. So that
there are three independent of equations.

We first consider solutions with $\omega_1\neq 0,\;\theta_1'\neq
0,\;\omega_2=0$ (or $\omega_1= 0,\;\theta_2'\neq 0,\;\omega_2\neq
0$). When $\theta_2'=0$, the equation of motion of $\theta_2$
implies that $\omega_2\sin\theta_2=0$. According to the previous
argument, classically it always indicates $\omega_2=0$, i.e.,
$\theta_2'=0,\Rightarrow\omega_2=0$. Those solutions, however, are
forbidden by inconsistency between the conformal
constraint~(\ref{3.2}) and the first equation in (\ref{3.4}). In
particular, the NS field plays crucial role in this assertion, as
we have mentioned.

Hence we should consider the multi-spin solutions with
$\omega_1\neq 0,\;\omega_2\neq 0$. It is convenient to rewrite
equations of motion~(\ref{3.4}) in terms of introducing
$\eta=\omega_1\cos\theta_1,\;\xi=\omega_2\cos\theta_2$ and
$z=\cos\sigma$,
\beq{3.17a}
(1-z^2)(\eta'-\xi')^2&=&
b^2[\omega_1^2+\omega_2^2-\eta^2-\xi^2+\frac{2}{3}
(\omega+\eta+\xi)^2+c]^2, \nn \\
(1-z^2)\eta'^2&=&\frac{1}{3}(\omega_1^2-\eta^2)(\eta^2-4\omega\eta
  -4\eta\xi+4F(z)+d_1), \nn \\
(1-z^2)\xi'^2&=& \frac{1}{3}(\omega_2^2-\xi^2)(\xi^2-4\omega\xi
 -4F(z)+d_2),
\eeq
where the ``prime'' denotes the differential to $z$,
$F(z)=\int\eta(z)\xi'(z)dz$, $d_1,\;d_2$ are integral constants
and $b,\;c$ are independent constants since they depend on
$r_*,\kappa,a$ and $R$. The appearance of the first equation
in~(\ref{3.17a}), which forbids some simpler solution such as
$\eta=\xi$, is entirely due to coupling between $r$ and $\theta$
and presence of NS B-field. Since $\eta$ and $\xi$ are required to
be periodic functions in region $\sigma\in [0,2\pi]$, the general
non-singular periodic solutions of $\eta$ and $\xi$ can be
expressed by polynomial of $z$,
\beq{3.17b}\eta=\sum_{n=0}^\infty a_nz^n,\hspace{1in}
\xi=\sum_{n=0}^\infty b_nz^n.
\eeq
Unfortunately, it is hard to answer whether the solution like
(\ref{3.17b}) is allowed or not for highly non-linear
equations~(\ref{3.17a}) since the above polynomial can not be
truncated at finite order.

\subsection{Folded solutions with $r'\neq 0$}
\label{sec33}

The equations of motion~(\ref{3.4}) in general is hard to solve
when $r'\neq 0$ due to coupling among $r$ and $\theta_i$.
Fortunately, these differential equations can be performed the
first integral if we set $\omega_1=\omega_2=0$. The solution reads
off
\beq{3.18}&&\theta_i'=\frac{c_i}{\sqrt{6(a\ln{(r/r_0)}+1)}},\nn \\
&&\frac{r'^2}{r^2}=\frac{\kappa^2r^2}{a\ln{(r/r_0)}+1}
-\frac{c_1^2+c_2^2}{a\ln{(r/r_0)}+1}-\frac{\omega^2}{9}.
\eeq
It is not hard to check that the above equations are consistent
with conformal constraints~(\ref{3.2}). The general solutions of
(\ref{3.18}) can not be expressed by elementary functions, so to
find the folded string solutions following from (\ref{3.18}) are
nontrivial task. For the simplest case with $a=0$, i.e., the
$AdS_5\times T^{1,1}$ background. It is straightforward to obtain
the following solution
\beq{3.19}
\theta_i&=&n_i\sigma, \hspace{0.6in}
r^2=r_0^2/\cos^2{\Omega\sigma},\nn \\
\kappa^2r_0^2&=&\Omega^2=\frac{1}{6}(n_1^2+n_2^2)+\frac{\omega^2}{9}.
\eeq
The above solution satisfies periodic boundary
condition~(\ref{3.5}) with $n_i$ integers and $\Omega$
half-integer. It is a circular string for nonzero $n_i$ and
folded one when $n_1=n_2=0$. The latter describes a closed string
spreading from $r_0$ to infinity and folded at
$r(\sigma=0)=r(\sigma=\pi)=r(\sigma=2\pi)=r_0$ and
$r(\sigma=\pi/2)=r(\sigma=3\pi/2)=\infty$. These solutions,
however, are ill-defined since they following from~(\ref{2.14})
yield infinite energy.

In general we may consider a first order differential equation
with the following form
\beq{3.22}r'^2=F(r)^2.
\eeq
If $F(r)$ is positive in region $0<r_1< r< r_2$, the function
$\upsilon(r)$ defined by integral
\beq{3.23}
\upsilon(r)=\int_{r_1}^r\frac{dr'}{F(r')}
\eeq
is single-valued in region $r_1< r< r_2,\;0< \upsilon <
\upsilon(r_2)$, so one can unambiguously define inverse function
of $\upsilon(r)$, $R(\upsilon)=\upsilon^{-1}(r)$ in this region.
Then we can construct a special solution of Eq.~(\ref{3.22}):
\beq{3.24}r=\left\{\begin{array}{ll}R(\sigma-\frac{2m\pi}{n}),
&\hspace{0.6in}{\rm if}\quad\sigma\in
(\frac{2m\pi}{n},\frac{(2m+1)\pi}{n}),\\
R(\frac{2(m+1)\pi}{n}-\sigma), &\hspace{0.6in}{\rm
if}\quad\sigma\in (\frac{(2m+1)\pi}{n},\frac{2(m+1)\pi}{n}),
\end{array}\right.
\eeq
with $m,\;n$ integers and $m=0,1,...,n-1$. The above solution
satisfies the boundary condition $r(\sigma+2\pi)=r(\sigma)$ and
has properties
\beq{3.21}
r(\frac{2m\pi}{n}+\sigma)=r(\sigma),\hspace{0.6in}
r(\frac{2(m+1)\pi}{n}-\sigma)=r(\sigma-\frac{2m\pi}{n}).
\eeq
The crucial point to admit this kind of solutions is because
Eq.~(\ref{3.22}) allows us to choose $r'=F(r)$ in the region
$2m\pi/n \leq \sigma\leq (2m+1)\pi/n$ and $r'=-F(r)$ in the region
$(2m+1)\pi/n \leq \sigma\leq 2(m+1)\pi/n$. Moreover, in order that
the solution~(\ref{3.24}) correctly describes the folded closed
string configuration, the turning points at $\sigma=k\pi/n$ must
be fixed points, i.e., $r'|_{\sigma=k\pi/n}=0$ or
$F(r(\sigma=k\pi/n))=0$. Without loss of generality, we may assume
$r_1,r_2$ are fixed points of Eq.~(\ref{3.22}). Then an additional
consistent condition,
\beq{3.25}\upsilon(r_2)=\frac{\pi}{n},
\eeq
has to be imposed.

Now let us turn to folded strings in deformed $AdS_5\times
T^{1,1}$. We may introduce new variable $\rho=r_0e^{-1/a}/r$ and
rewrite the second equation in~(\ref{3.18}) as follows
\beq{3.26}\rho'^2=\frac{\kappa^2 r_0^2e^{-2/a}}{a}
\frac{\bar{c}^2\rho^2-1-\bar{\omega}^2\rho^2\ln{\rho}}{\ln{\rho}},
\eeq
where
\beq{3.27}\bar{c}^2=\frac{c_1^2+c_2^2}{\kappa^2r_0^2e^{-2/a}},
\hspace{1in}
\bar{\omega}^2=\frac{a\omega^2}{9\kappa^2r_0^2e^{-2/a}}.
\eeq
In the physical region, $\rho\in [0,e^{-1/4})$, the
equation~(\ref{3.26}) can only have two possible fixed points:
$\rho=0$ and $\rho=\rho_c$ with $\rho_c$ solution of algebraic
equation
\beq{3.28}\bar{c}^2\rho^2-1-\bar{\omega}^2\rho^2\ln{\rho}=0.
\eeq
In the region $\rho\in (0,\rho_c)$ it is possible to choose
$\bar{c}$ and $\bar{\omega}$ to ensure r.h.s. of~(\ref{3.26}) to
be positive. Then if condition following from (\ref{3.27}),
\beq{3.29}\int_{0}^{\rho_c}d\rho\sqrt{\frac{\ln{\rho}}
{\bar{c}^2\rho^2-1-\bar{\omega}^2\rho^2\ln{\rho}}} =\frac{\kappa
r_0e^{-1/a}}{\sqrt{a}}\frac{\pi}{n},
\eeq
is satisfied, we may construct a solution like (\ref{3.24}). This
solution describes a folded closed string spreading from $\rho=0$
to $\rho_c$ (or from $r_c$ to infinity). Indeed it is not hard to
prove that the integral in~(\ref{3.29}) is convergent. For
example, if we choose $\bar{c}=1,\;\bar{\omega}=5$ and $n=4$, we
obtain $\rho_c=0.1416$ and $\kappa r_0e^{-1/a}/\sqrt{a}=0.49$.

It is interesting that the equation~(\ref{3.18}) admits static
configuration ($\omega=0$), which is not admitted for $AdS_5\times
T^{1,1}$.  Explicitly, for $\omega=0$ we have
\beq{3.30}r=\frac{\kappa^{-1}\sqrt{c_1^2+c_2^2}}
{|\cos\sqrt{\theta_1^2+\theta_2^2}|}.
\eeq
By means of the boundary conditions listed in~(\ref{3.5}) it is
easy to check that the configuration describes a circular closed
string when
\beq{3.31}m_1\theta_1=m_2\theta_2,\hspace{1in}
\frac{m_1}{m_2}\sqrt{m_1^2+m_2^2}\in\mathbb{Z},
\eeq
with winding number $m_1,\;m_2$ integers. For other cases we have
to set $m_1=m_2=0$ so that it describes a folded string spreading
from $r_c=\sqrt{c_1^2+c_2^2}/\kappa$ to infinity. However, it is
tilted folded configuration comparing with one on $AdS_5\times
T^{1,1}$. Namely both of $\theta_i$ and $r$ acquire
$\sigma$-dependence. Those static string configurations yield
infinite energy
\beq{3.32}E_s=\frac{1}{\ep}+...,
\eeq
with $\ep$ a cut-off. Moreover, the spinning folded string
configurations found previously yield infinity energy too,
\beq{3.33}E_{spin}=E_s+E_0+F(J),\hspace{0.5in}F(J)>0,
\eeq
with $E_0$ finite constant. Although the infinite energy implies
that those configurations are not well-defined, it may make sense
to ask whether energy/charge relation ${\cal
E}=E_{spin}-E_s=E_0+F(J)$ can be reproduced from dual SYM. If
true, the usual relation ${\cal E}\propto J$ is modified, as we
expected$^3$.\footnotetext[3]{We do not know whether $F(J)\propto
J$ at large $J$ limit is still kept. It need get the explicit
expression of the solutions.}

\section{Type B solution}
\setcounter{equation}{0}

Here the word ``type B'' means that in this section we shall use a
ansatz in which
\beq{4.1}\psi'=\dot{\theta}_1=\dot{\theta}_1=0,
\eeq
but $\phi_1$ and $\phi_2$ acquire both of $\tau$- and
$\sigma$-dependence. Then following from Eq.~(\ref{2.9}) the
equation of motion of $\psi$ reads off
\beq{4.2}3\ddot{\psi}=-\sum_{i=1}^2\frac{3+\cos^2\theta_i}{\sin\theta_i}
\theta_i'\phi_i'+2\cot\theta_1\cot\theta_2
(\sin\theta_2\;\theta_1'\phi_2'+\sin\theta_1\;\theta_2'\phi_1').
\eeq
The $\sigma$-independence of $\psi$ requires the r.h.s of the
above equation is independent of $\sigma$ too. The simplest
solution is then
\beq{4.3}\theta_1'\phi_1'=\theta_2'\phi_1'=\theta_1'\phi_2'
=\theta_2'\phi_2'=0,\hspace{1in}\psi=\omega\tau.
\eeq
Since we hope that at least one of $\phi_i'$ does not vanish, we
have $\theta_1'=\theta_2'=0$. The conformal constraints in
(\ref{2.10}) reduce to
\beq{4.4}
&&\frac{k^2}{r^2h(r)}=\frac{r'^2}{r^2}+
\frac{1}{6}\sum_{i=1}^2\sin^2\theta_i(\dot{\phi}_i^2+\phi_i'^2)
+\frac{1}{9}(\omega+\dot{\phi}_1\cos\theta_1+\dot{\phi}_2\cos\theta_2)^2
\nn \\
&&\quad\quad\quad\quad\quad
+\frac{1}{9}(\phi_1'\cos\theta_1+\phi_2'\cos\theta_2)^2,
\nn \\
&&\sum_{i=1}^2(1-\frac{1}{3}\cos^2\theta_i)\dot{\phi}_i\phi_i'
+\frac{2}{3}\cos\theta_1\cos\theta_2(\dot{\phi}_1\phi_2'
+\dot{\phi}_2\phi_1')=0.
\eeq
Using those constraints and the above ansatz in equations of
motion, one has
\beq{4.5}
&&\frac{r''}{r}-\frac{r'^2}{r^2}-\frac{\kappa^2}{r^2h}
(1-\frac{a/2}{1+a\ln{(r/r_0)}})\nn \\
&&\quad\quad\quad-\frac{a/12}
{1+a\ln{(r/r_0)}}\left\{\sum_{i=1}^2\phi_i'^2\sin^2\theta_i
+\frac{2}{3}(\phi_1'\cos\theta_1+\phi_2'\cos\theta_2)^2\right\}=0,
\nn \\
&&2\omega\dot{\phi}_1-(\dot{\phi}_1^2-\phi_1'^2)
\cos\theta_1+2(\dot{\phi}_1\dot{\phi}_2-\phi_1'\phi_2')\cos\theta_2
+\frac{3R^2\sqrt{a}}{r^3\sqrt{2h}}r'\dot{\phi}_1=0, \nn \\
&&2\omega\dot{\phi}_2-(\dot{\phi}_2^2-\phi_2'^2)
\cos\theta_2+2(\dot{\phi}_1\dot{\phi}_2-\phi_1'\phi_2')\cos\theta_1
-\frac{3R^2\sqrt{a}}{r^3\sqrt{2h}}r'\dot{\phi}_2=0, \nn \\
&&-\ddot{\phi}_i+\phi_i''+\frac{a\;\phi_i'}{1+a\ln{(r/r_0)}}
\frac{r'}{2r}=0,\hspace{0.5in}(i=1,2).
\eeq
Here we used the fact that $\sin\theta_i=0$ corresponds to string
shrink to a point at south or north-pole of $S^2$. It is
uninterested by us, i.e., we shall always assume that
$\sin\theta_i\neq 0$ in rest part of this paper. It should be
noticed that, in equations of $r$ and $\phi_i$, the terms
involving NS field vanish. Moreover, the equation~(\ref{4.5})
should be supplemented by boundary condition
$\phi_i(\sigma+2\pi)=\phi_i(\sigma)+2m_i\pi$ with $m_i$ integers.

\subsection{Circular solutions}

Let us first study the circular solutions by imposing $r'=0$. Here
equations of motion of $\phi_i$ reduce to one of free fields on
world-sheet. It has the general solution
\beq{4.6}\phi_i=\omega_i\tau+m_i\sigma
+\sum_n\{a_{in}\cos{n(\tau+\sigma)}+b_{in}\cos{n(\tau-\sigma)}\},
\eeq
with $m,\;n$ integers. Meanwhile, we notice that the last term in
the first equation of (\ref{4.5}) has to be constant. Substituting
the solution~(\ref{4.6}) into this equation one has
$a_{in}=b_{in}=0$. Furthermore, the consistence between the first
equation of motion in (\ref{4.5}) and the first conformal
constraint in (\ref{4.4}) implies
\beq{4.7}r=r_0\exp{\left(\frac{\Omega^2}{2(\Omega^2+M^2)}-\frac{1}{a}
\right)},
\eeq
where
\beq{4.8}
\Omega^2&=&\sum_{i=1}^2\omega_i^2\sin^2\theta_i
+\frac{2}{3}(\omega+\omega_1\cos\theta_1
+\omega_2\cos\theta_2)^2, \nn \\
M^2&=&\sum_{i=1}^2m_i^2\sin^2\theta_i
+\frac{2}{3}(m_1\cos\theta_1+m_2\cos\theta_2)^2.
\eeq
The ``physical'' condition $N_{\rm eff}>0$ further imposes that
$\Omega^2>M^2$. The above results have been enough to determine
the classical energy following from eqs.~(\ref{2.14}) and
(\ref{4.4}):
\beq{4.8a}E=\frac{\sqrt{2}r_0}{\sqrt{3}\alpha'}e^{-1/a}
\sqrt{\Omega^2+M^2}
\exp{\left(\frac{\Omega^2}{2(\Omega^2+M^2)}-\frac{1}{a} \right)}.
\eeq
It exhibits an interesting exponential factor, like one for
circular string in ``type A'' solutions. The task to explore the
relation between $E$ and various spins then reduces to find the
relation between $\Omega,\;M$ and spins. For all of the following
special solutions we shall show that we can always define a
convenient ``angular momentum'' satisfying
\beq{4.8b}{\cal J}^2=\frac{\Omega^4}{2(\Omega^2+M^2)}.
\eeq
Then at large charge limit ${\cal J}\to\infty$, the classical
energy can be expand regularly
\beq{4.8c}E=\frac{2r_0}{\sqrt{3}\alpha'}e^{\frac{1}{2}-\frac{1}{a}}{\cal
J}(1+\frac{M^2}{{\cal J}^2}+...) .
\eeq
In order to find the explicit solutions, we shall solve the
remainder equations of motion that now reduces to the linear
algebraic equations on $\cos\theta_i$,
\beq{4.9}\left\{\begin{array}{ll}
(\omega_1^2-m_1^2)\cos\theta_1-2(\omega_1\omega_2-m_1m_2)
\cos\theta_2=2\omega\omega_1, & \\
(\omega_2^2-m_2^2)\cos\theta_2-2(\omega_1\omega_2-m_1m_2)
\cos\theta_1=2\omega\omega_2,
\end{array}\right.
\eeq
supplemented by the second conformal constraint in (\ref{4.4}).
The equations in (\ref{4.9}) has several different solutions:
\begin{itemize}
\item[-] $\omega\omega_1=\omega\omega_2=0$.
\end{itemize}

The equation~(\ref{4.9}) in this case has trivial solution
$\cos\theta_1=\cos\theta_2=0$ unless
\beq{4.10}\omega_1\omega_2=m_1m_2,\;\omega_1m_2=\omega_2m_1,
\quad\Rightarrow\quad\omega_i^2=m_i^2.
\eeq
Using the above equation to the second conformal constraint in
(\ref{4.4}) one has $\omega_1=\omega_2$. If $\omega=0$, therefore,
the nontrivial spinning configuration is presented only for the
trivial solution $\cos\theta_1=\cos\theta_2=0$.
\begin{enumerate}
\item Two-spin solution: $\omega=0,\;\cos\theta_1=\cos\theta_2=0$.

The second conformal constraint in (\ref{4.4}) indicates
$\omega_1m_1+\omega_2m_2=0$, while equation~(\ref{4.8}) reduces to
$$\Omega^2=\omega_1^2+\omega_2^2,\hspace{0.5in}M^2=m_1^2+m_2^2.$$
The conserved charges following from eqs.~(\ref{2.13}),
(\ref{2.14}) and (\ref{4.4}) read off
\beq{4.11}
J_A&=&\frac{R^2}{3\alpha'}\omega_1\sqrt{\frac{a\Omega^2}{2(\Omega^2+M^2)}},
\hspace{0.5in} J_B=J_A(\omega_1\leftrightarrow\omega_2),
\hspace{0.5in}J_R=0, \nn \\ {\cal
J}&=&\frac{3\alpha'}{R^2\sqrt{a}}\sqrt{J_A^2+J_B^2}.
\eeq
Its dual operators in SYM should be like
\beq{4.12}{\rm Tr}(A_1A_2^\dag)^{J_A}(B_1B_2^\dag)^{J_B}.
\eeq
It is out of holomorphic sector.
\item Single-spin solution: $\omega\neq 0,\;\omega_1=\omega_2=0$.

In this case the equation~(\ref{4.9}) admits two solutions that
correspond to different operators in dual field theory. One of
them is $\cos\theta_1=\cos\theta_2=0$ which leads the conserved
charges
\beq{4.13}J_A=J_B=0,\hspace{0.6in}J_R=
\frac{R^2}{9\alpha'}\sqrt{\frac{2a}{3}}{\cal
J}=\frac{2R^2}{9\alpha'}
\omega\sqrt{\frac{a\omega^2}{2\omega^2+3M^2}}.
\eeq
For even $J$, its dual operators in the holomorphic sector have
the form
\beq{4.14}{\rm Tr}(A_1B_1)^{J/2}(A_2B_2)^{J/2}.
\eeq
If $J$ is odd, they should be no longer to be holomorphic
operators. Meanwhile, another solution $m_1=0,\;\cos\theta_2=0,
\;\cos\theta_1={\rm arbitrary\;\; constant}$ (or
$m_2=0,\;\cos\theta_1=0, \;\cos\theta_2$ arbitrary) leads to
\beq{4.15}J_R&=&\frac{R^2}{9\alpha'}\sqrt{\frac{2a}{3}}{\cal
J}=\frac{2R^2}{9\alpha'}
\omega\sqrt{\frac{a\omega^2}{2\omega^2+3M^2}},\nn \\
J_A&=&J\cos\theta_1,\hspace{1in}J_B=0.
\eeq
Its dual operators must be out of holomorphic sector. In addition,
if $\cos\theta_1\neq n/J_R$ with $n,\;J_R$ integers, the $U(1)_A$
is anomalous.
\end{enumerate}
\begin{itemize}
\item[-] $\omega\omega_1\neq 0$ and/or $\omega\omega_2\neq 0$.
\end{itemize}

For this assumption the solutions of equation~(\ref{4.9}) together
with the second conformal constraint in (\ref{4.4}) in general
lead to tedious expressions of angular momenta and classical
energy. For the sake of convenience, we consider an example of
two-spin solution with $\omega_2=0$ that leads the solution of
(\ref{4.9}) as follows
\beq{4.17}\cos\theta_1=\frac{2\omega\omega_1}{\omega_1^2+3m_1},
\hspace{0.5in}\cos\theta_2=\frac{4\omega\omega_1m_1}
{m_2(\omega_1^2+3m_1)}.
\eeq
Substituting the above solution into the second conformal
constraint in (\ref{4.4}) one has
\beq{4.18}m_1=0,\quad\to\quad\cos
\theta_1=\frac{2\omega}{\omega_1},\hspace{0.3in}\cos\theta_2=0.
\eeq
It is then straightforward to write the explicit expressions for
angular momenta,
\beq{4.19}J_R&=&\frac{2R^2}{3\alpha'}
\omega\sqrt{\frac{a\Omega^2}{2(\Omega^2+M^2)}},\hspace{0.5in}
J_A=\frac{\omega_1}{2\omega}J_R,\hspace{0.5in}J_B=0, \nn \\
{\cal J}&=&\frac{3\alpha'}{R^2\sqrt{a}}\sqrt{J_A^2+J_R^2/2}.
\eeq
Its dual operators again must be out of holomorphic sector.

\subsection{Folded solution with $r'\neq 0$}

From eq.~(\ref{4.5}) the most general solution of $\phi_i$ in this
case has the following form:
\beq{4.20}\phi_i(\tau,\sigma)=\eta_i\tau^2+
\omega_i\tau+T_i(\tau)S_i(\sigma).
\eeq
Recalling that $r$ is independent of $\tau$, the second and the
third equations in~(\ref{4.5}) indicates that $\eta_i=0$, while
the first equation in~(\ref{4.5}) implies $T_i$ to be constants.
Hence it is straightforward to obtain
\beq{4.21}S_i'=\frac{c_i}{\sqrt{1+a\ln{(r/r_0)}}},
\eeq
with $c_i$ constants. Substituting Eqs.~(\ref{4.20}) and
(\ref{4.21}) into the second and the third equations
in~(\ref{4.5}) we can easily to check that they are inconsistent
with the first conformal constraint in~(\ref{4.4}) unless
$\omega_1=\omega_2=0$. In other words, there is only single-spin
``type B'' solution when $r'\neq 0$. It is different from one in
``type A'' solutions where we can not claim the same conclusion
although it is hard to solve corresponding equations of motion
with two or three nonzero spins.

In terms of the above results, there is a remainder constraint,
$c_1\cos\theta_1=c_2\cos\theta_2=0$, following from
Eq.~(\ref{4.5}). Finally to solve the equations of motion of $r$
we get nothing but just Eq.~(\ref{3.18}). Hence following the same
process in section~(\ref{sec33}) we can show that there are folded
string solutions with infinite energy.

\section{Summary and Discussions}
\setcounter{equation}{0}

We have studied various spinning closed string configurations on
Klebanov-Tseytlin background. This background is described by a
logarithmically deformed $AdS_5\times T^{1,1}$ geometry induced by
nontrivial Neveu-Schwarz flux and is dual to a nonconformal
$SU(N+M)\times SU(N)$ $\CN{1}$ SYM. We studied spinning string
solutions with two types of different ansatzs. They describe the
classical closed strings rotating on $T^{1,1}$ part of KT
background. Explicit expressions for point-like solutions and some
circular solutions were found and the energy/charge relation for
these solutions were obtained. The folded string solutions along
the radial direction was discussed. These folded solutions will in
general break usual energy/charge relation $E\propto J$ at large
$J$ limit. The dual operators in gauge theory were constructed for
diverse solutions.

\subsection{Comparing with solutions in $AdS_5\times T^{1,1}$
background}

To manifest how presence of NS field and deformation of geometry
change the integrability of original $AdS_5\times T^{1,1}$
background, it is worth to compare the solutions obtained in this
present paper with those obtained in $AdS_5\times T^{1,1}$
geometry. It is obvious that the KT geometry reduces to
$AdS_5\times T^{1,1}$ at the limit $a\to 0$, where the $AdS$
metric is given by Poincar$\acute{\rm e}$ coordinate. If none of
Poincar$\acute{\rm e}$ coordinates of $AdS_5$ acquiring
$\sigma$-dependence, there is no finite energy the spinning string
solutions $T^{1,1}$ with static gauge $x^0=\tau$ unless the radius
of $AdS_5$ goes to infinity. This
conclusion$^4$\footnotetext[4]{The assertion is not right for
global coordinate of $AdS_5$. It is known that there are the
spinning string solutions in $S^5$ or $T^{1,1}$ at the center of
$AdS_5$\cite{Tse03} without requirement of the limit $R\to\infty$.
The key point is that the static gauge in global coordinate is no
longer to be static one in Poincar$\acute{\rm e}$ coordinate. In
other words, both of $x^0$ and $r$ in Poincar$\acute{\rm e}$
coordinate acquire not only $\tau$-dependence, but also
$\sigma$-dependence. In this present paper we insist on using the
Poincar$\acute{\rm e}$ coordinate because the radial coordinate
has manifest physical meaning in dual quantum field theory.}
follows from Eqs.~(\ref{2.14}) and (\ref{3.2}) that indicates that
$E\propto r$ while Eq.~(\ref{3.4}) requires $r^2/R^4\to 0$ when
$r'=0,\;a\to 0$. The KT background, in contrast, allows the finite
energy spinning string configurations in $T^{1,1}$ without
requirement to take the limit $R\to\infty$. These solutions,
however, locate at special region along the radius of $AdS_5$: The
point-like strings locate at
$r=r_*=r_0\exp{(\frac{1}{2}-\frac{1}{a})}$. While the circular
strings are presented in the region $\td{r}_s<r\leq r_*$.

From perspective of world-sheet sigma model, the KT background
bring some interesting features in contrast to $AdS_5\times
T^{1,1}$ background: 1) The equation of motion of radial
coordinate $r$ does not decouple with those of $T^{1,1}$
coordinates. This feature induces extra constraints to usual
spinning string solutions on $AdS_5\times T^{1,1}$ background. So
that some solutions are forbidden. 2) The NS $B$-field is involved
in equations of motion. In fact, the NS field plays trivial role
in all of solutions presented in this present paper. So far all
attempts to find the solutions in which the NS field plays
nontrivial role are failed. In particular, we have shown that
those simpler circular solutions are forbidden by NS field role.
For solutions with $r'\neq 0$, it is easy to check that the NS
field plays nontrivial role only if $\omega_i\neq 0$ and
$\theta_i'\neq 0,\;\theta_1\neq\theta_2$ following from
Eqs.~(\ref{3.2}) and (\ref{3.4}). Seemingly it is impossible to
find analytic solutions satisfying all of these requirements. In
other words, the presence of NS field should break the
integrability of $AdS_5\times T^{1,1}$ background.

\subsection{Challenges in dual gauge theory}

We expect that the various solutions found in the present paper
enable us to precisely test general gauge/string duality, like
those for AdS/CFT correspondence. Many interesting features of
these solutions, however, bring new challenges to understand them
from perspective of dual quantum field theory. For example, how to
obtain an integrable spin chain from SYM, and how to construct the
dual operators of folded strings?

We have mentioned that in our solutions, the point-like strings
locate at $r=r_*=r_0\exp{(\frac{1}{2}-\frac{1}{a})}$ and the
circular strings are presented in the region $\td{r}_s<r\leq r_*$.
It implies that $r=r_*$ plays special roles in these solutions.
Since the flow along radial direction of background in string
theory is known to be dual to renormalization group flow in
quantum field theory, it indicates that there is a special energy
scale in quantum field theory where a sort of phase transition
occurs. It is not necessary that this scale is at IR since the
value of $r_*$ depends on $a\sim g_sM^2/N$. Although in KT
background one has $M/N\ll 1$, it is possible to impose a large
$g_sM$ such that $a$ has finite value. Moreover, the phase
transition does not correspond to the breaking of chiral symmetry
associating with gauge group $SU(N+M)\times SU(N)$ revealed in
\cite{KS00}. Here it should be associated with the global symmetry
$SU(2)\times SU(2)$.

In all of our solutions, the charges $J\propto
R^2\sqrt{a}\omega/\alpha'\sim g_sM\omega$. Hence the real
parameter controlling regular expansion at double limit is
$\lambda_{_M}^2/J^2$ where $\lambda_{_M}=g_sM$ is 't Hooft
couplings corresponding to $SU(M)$ group . While from the point of
view of dual SYM, there are two 't Hooft couplings
$\lambda_1=g_1^2(N+M)$ and $\lambda_2=g_2^2N$ in which
$g_1^{-2}+g_2^{-2}\sim g_s^{-1}\sim$constant, but
$g_1^{-2}-g_2^{-2}$ flows logarithmically under renormalization
group. These known results yield
$$\lambda_{_M}\;\sim\;\frac{\lambda_1\lambda_2}
{\lambda_1N/M+\lambda_2(1+N/M)}.$$ It is puzzled since seemingly
it is out of one-loop corrections of quantum field theory.

In several solutions we have obtained an implication that the
$U(1)_A$ and/or $U(1)_B$ presented in table 1 should be anomalous.
Therefore another question is how to explore this phenomena from
dual SYM. In addition, in ref.\cite{BCMZ04} authors studied the
hadron dynamics in Witten model in terms of semiclassical
solutions in dual string theory. The similar studies are expected
to be performed for the KT background and its dual gauge theory.

\acknowledgments{It is pleasure to thank A.A. Tesytlin who pointed
out a mistake in conformal constraint. This work is partly
supported by the NSF of China, Grant No. 10305017, and through
USTC ICTS by grants from the Chinese Academy of Science and a
grant from NSFC of China.}

\end{document}